\documentclass[aps, prb, reprint, superscriptaddress, showpacs]{revtex4-1}

\usepackage{amsmath}
\usepackage{graphicx}
\usepackage{bm}
\usepackage{amssymb}
\usepackage{subfigure}
\usepackage[colorlinks, linkcolor=blue, citecolor=blue]{hyperref}
\let\sss= \scriptscriptstyle

\begin{document}

  \title{Exploring quantum phase transitions by the cross derivative of the ground state energy}

    \author{H. Y. Wu}
    \affiliation{School of Physics and Electronics, Hunan University, Changsha 410082, China}
    \affiliation{Ningbo Institute of Materials Technology and Engineering, CAS, Ningbo 315201, China}
    \author{Yu-Chin Tzeng}
    \affiliation{Physics Division, National Center for Theoretical Sciences, Taipei 10617, Taiwan}
    \affiliation{Center for Theoretical and Computational Physics, National Yang Ming Chiao Tung University, Hsinchu, 300093 Taiwan}
    \author{Z. Y. Xie}
      \email{qingtaoxie@ruc.edu.cn}
      \affiliation{Department of Physics, Renmin University of China, Beijing 100872, China}
    \author{K. Ji}
      \email{kji@shnu.edu.cn}
      \affiliation{Department of Physics, Shanghai Normal University, Shanghai 200234, China}
    \author{J. F. Yu}
      \email{yujifeng@hnu.edu.cn}
      \affiliation{Department of Applied Physics, Hunan University, Changsha 410082, China}


\begin{abstract}
In this work, the cross derivative of the Gibbs free energy, initially proposed for phase transitions in classical spin models [Phys. Rev. B 101, 165123 (2020)], is extended for quantum systems. We take the spin-1 {\textit{XXZ}} chain with anisotropies as an example to demonstrate its effectiveness and convenience for the Gaussian-type quantum phase transitions therein. These higher-order transitions are very challenging to determine by conventional methods. From the cross derivative with respect to the two anisotropic strengths, a single valley structure is observed clearly in each system size. The finite-size extrapolation of the valley depth shows a perfect logarithmic divergence, signaling the onset of a phase transition.
Meanwhile, the critical point and the critical exponent for the correlation length are obtained by a power-law fitting of the valley location in each size. The results are well consistent with the best estimations in the literature. Its application to other quantum systems with continuous phase transitions is also discussed briefly.
\end{abstract}

 \maketitle

\section{Introduction}\label{Introduction}

Exploring novel phases of matter and phase transitions has always been one of the central topics in statistical and condensed matter physics.
It has greatly enriched our understanding of matter phases and attracted much attention and effort since the discovery of the topological phases and phase transitions \cite{KT1,KT2,Haldane1,Haldane2,Haldane3} beyond Landau's symmetry-breaking theory.
Previously, we have proposed and demonstrated that the cross derivative of the Gibbs free energy is efficient and convenient for detecting various phase transitions in classical spin models\cite{YuPRB2020}, no matter whether a transition is conventional or exotic with topological excitations.
Its success inspires us to extend its applicability to quantum cases, wherein the driving forces of a phase transition are much more diversified. Accordinglly, the Gibbs free energy is reduced to the ground state energy, and the temperature fluctuation is replaced by the frustration effect between multiple competing interactions.

Usually, the introduction of competition into a quantum system raises complexity and difficulty in identifying phase transitions and the phase diagram. In particular, the transitions higher than 2nd-order are tough to determine precisely.
One typical example is the spin-1 {\textit{XXZ}} chain with the single-ion anisotropies, where the 3rd- and 5th-order Gaussian-type quantum phase transitions have been suggested for different parameters \cite{ZengPRB2017, Wang2011,ZengPRA22008,BoschiEPJB2003}. 
For these transitions, the conventional 2nd-order differential of the ground state energy doesn't work as usual, and the fidelity susceptibility is not necessarily applicable.\cite{ZengPRA2008}

Because of its rich phase diagram, this model has been widely utilized to study effective one-dimensional spin-1 magnetic materials, as reviewed in Ref.\ [\onlinecite{Spin1:compounds}]. It also serves as a testing ground for sophisticated numerical methods. Recently, the tunable single-ion anisotropic effects have been realized experimentally in spin-1 models with ultra-cold atoms\cite{Woo},  and in the compound $[\mathrm{Ni}(\mathrm{HF}_2)(3\mbox{-}\mathrm{Clpyradine})_4]\mathrm{BF}_4$ with inelastic neutron scattering by pressure variance\cite{NBCT}.
There are more details about the anisotropic effects and the phase diagram of this system in Refs.\ [\onlinecite{ZengPRB2017,RenPRA2018,Wang2011,JohnstonPRB,albuqPRB,GHLiuPB,ChenPRB2003}]. 
Due to the inefficiency of the conventional differential tool, it is usually studied with unique tactics such as entanglement entropy, fidelity susceptibility, etc.\cite{ZengPRB2017, Wang2011, LyraPRB, arXiv2207_05052, RenPRA2018, ZengPRA22008}.

In this work, we use this model as an example to demonstrate that, by making use of competing interactions, the cross derivative captures the essential characteristics of the phase transitions. By tuning the uniaxial anisotropy strength $D$ and the rhombic one $E$, we first investigate the 3rd-order Gaussian-type transitions in this model when $J_z=1$. In the $(D, E)$-plane, the critical point between the Haldane phase and the Large-$D$ phase is determined at $(0.9687, 0)$, and the Haldane-Large-$E$ critical point locates at $(-0.4862, 0.4862)$. Both are well consistent with previous predictions\cite{ZengPRB2017, Wang2011, ChenPRB2003}. Moreover, the critical exponent for the correlation length is obtained simultaneously. We also study the more difficult 5th-order Gaussian-type Haldane-Large-$D$ transition when $J_z=0.5$. The critical point is estimated precisely at (0.6197, 0), which agrees well with best estimations in the literature\cite{Wang2011,ChenPRB2003,ZengPRA22008}. Thus briefly, our cross derivative method provides a convenient, efficient, and universal tool to detect phase transitions not only in classical spin models\cite{YuPRB2020} but also in quantum systems, whether a transition is conventional or exotic with higher order.

The rest of the paper is organized as follows. Sec.\ \ref{ModelMethod} introduces the model and the method employed in our work.
The results are presented in Sec.\ \ref{Results}. Finally, Sec.\ \ref{Summary} gives a summary and discussion.


\section{Model and Method}\label{ModelMethod}

In this work, we study the spin-1 $XXZ$ chain with anisotropies, whose Hamiltonian reads as
  \begin{align}
    H=\sum_{i=1}^{L} ({S}^x_i  {S}^x_{i+1}+{S}^y_i  S^y_{i+1} + J_zS_i^zS_{i+1}^z) \nonumber\\
    + D\sum_{i=1}^L (S_i^z)^2 + E\sum_{i=1}^L [(S_i^x)^2-(S_i^y)^2], \label{eq1}
  \end{align}
where $S_i^\alpha (\alpha=x, y, z)$ are spin-1 operators on the $i$-th site, and $L$ is the length of the spin chain. The strength of the Heisenberg exchange interaction in $xy$-plane is set to $1$ for convenience. $J_z$ is the anisotropy strength in $z$-direction with respect to $xy$-plane. In this work, we only consider two special cases, i.e. $J_z=1$ and $J_z=0.5$. $D$ and $E$ are the strengths of the uniaxial and rhombic single-ion anisotropies, respectively. 

To calculate the ground state energy, we employed the density matrix renormalization group (DMRG) method \cite{White1,White2, Schollwock1}, by using ladder scheme and encoding parity symmetry\cite{ZengPRB2012} with a periodic boundary condition.  Regarding the accuracy with respect to the bond-dimension $m$, we follow the strategy of Ref.\ [\onlinecite{ZengPRB2012}] to choose big enough $m$ to make sure the truncation error is smaller than $10^{-9}$ for each system size, thus to guarantee the obtained energy is precise enough for further differential computation. Practically, $m$ also increases gradually with system size, i.e., $m=700$ for $L=40$, and $m=1200$ for $L=80$. The obtained result is also a good reflection of the computing precision. For illustration, we usually adopt an intermediate system size with $L=40$. Afterward, a finite size extrapolation to the thermodynamic limit is performed.

  \begin{figure}[htbp]
    \begin{center}
      \includegraphics[width=0.48\textwidth,clip,angle=0]{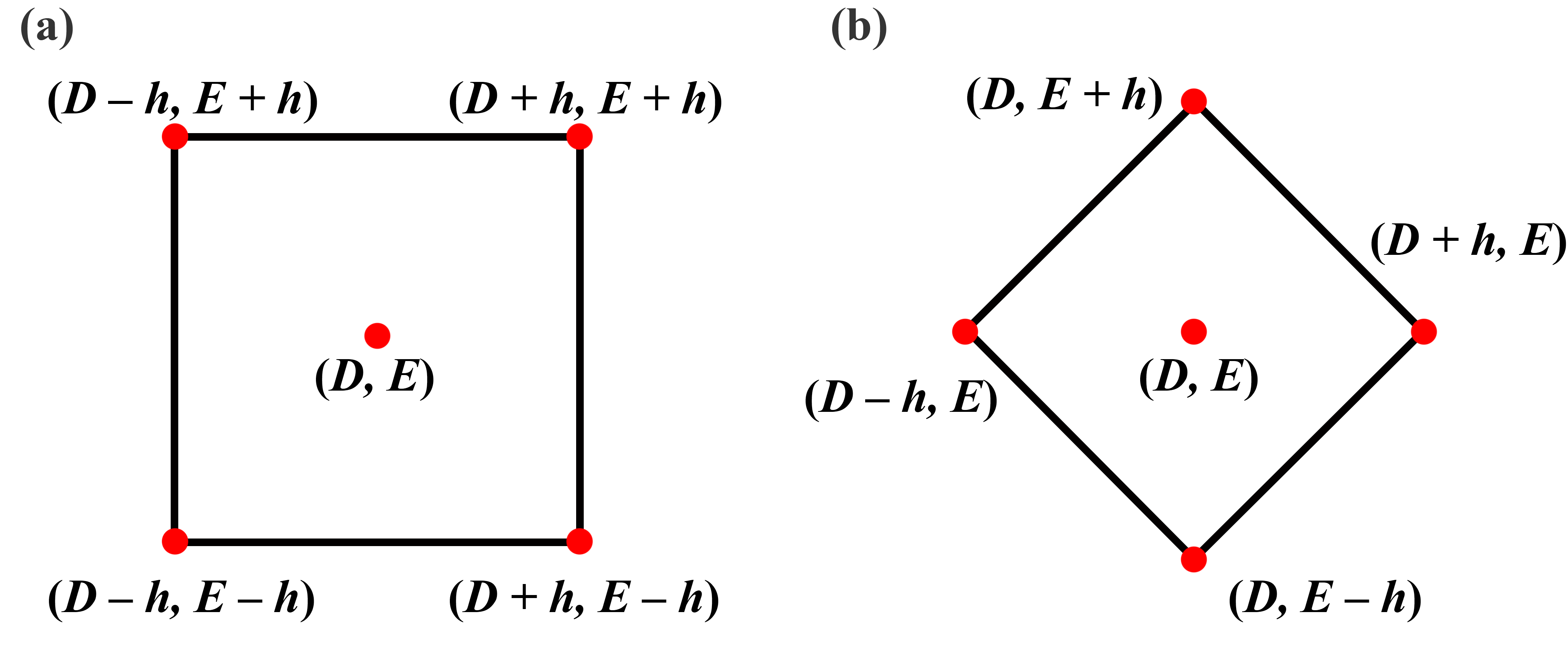}
      \caption{\label{CD_definition} (Color online) Two ways of calculating the cross derivative: (a) the normal one; (b) the rotated one.}
    \end{center}
  \end{figure}

The quantity proposed to detect the quantum phase transitions, is the cross derivative of the ground state energy density ($\epsilon=\langle H \rangle / L$) with respect to the anisotropic strengths $D$ and $E$,  i.e., $\partial^2\epsilon/\partial{D}\partial{E}$. 
Mathematically, to get complete information of a 3-dimensional surface $f(x, y, z)=0$ at a given point $(x, y, z)$, one needs in principle not only the curvatures in $x$- and $y$-directions but also the twist, namely, 
  \begin{equation}
    \frac{\partial^2}{\partial{t^2}}=\alpha^2\frac{\partial^2}{\partial{x^2}}+2\alpha\beta\frac{\partial^2}{\partial{x}\partial{y}}+\beta^2\frac{\partial^2}{\partial{y^2}}, \label{eq2}
\end{equation}
where $\partial/\partial{t}\equiv\alpha\partial/\partial{x}+\beta\partial/\partial{y}$ is the slope operator in an arbitrary direction. An equal weight is chosen for simplicity, i.e., $\alpha=\beta$. As demonstrated below, this cross derivative contains the contributions from both orthogonal directions, and then is able to detect quantum phase transitions, especially those higher-order ones, for which the curvature in either principal direction is inadequate.

To obtain this quantity at any point in the $(D, E)$-plane, we apply a normal central differential formula as
  \begin{equation}
    \frac{\partial^2\epsilon}{\partial{D}\partial{E}}=\frac{\epsilon_{\sss{D+h,E+h}}-\epsilon_{\sss{{D+h,E-h}}}-\epsilon_{\sss{{D-h,E+h}}}+\epsilon_{\sss{{D-h,E-h}}}}{4h^2},\label{eq3}
  \end{equation}
where $h$ is set to $10^{-3}$. The error induced by this formula is of order $O(h^2)$. Since the Hamiltonian is symmetric between $S^x$ and $S^y$, $\epsilon_{\sss{D,E}}$ and $\epsilon_{\sss{D,-E}}$ are equal, and thus the above cross derivative on the line $E=0$ is zero. In this situation, we instead use a rotated one as
  \begin{align}
    \frac{\partial^2{\epsilon}}{\partial{X}\partial{Y}} & \equiv\frac{ \epsilon_{\sss{D, E+h}}-\epsilon_{\sss{D+h, E}}-\epsilon_{\sss{D-h, E}}+\epsilon_{\sss{D, E-h}} } {2h^2}\nonumber\\
   &\sim \frac{\partial^2\epsilon}{\partial{E^2}}-\frac{\partial^2\epsilon}{\partial{D^2}},\label{eq4}
  \end{align}
where the parameters $X=(D+E)/\sqrt{2}$, and $Y=(D-E)/\sqrt{2}$. It is equivalent to rotating the coordinate frame clockwise by $\pi/4$ , as schematically shown in Fig.\ \ref{CD_definition}.

\section{Results}\label{Results}

\begin{figure}[htbp]
    \begin{center}
      \includegraphics[width=0.45\textwidth,clip,angle=0]{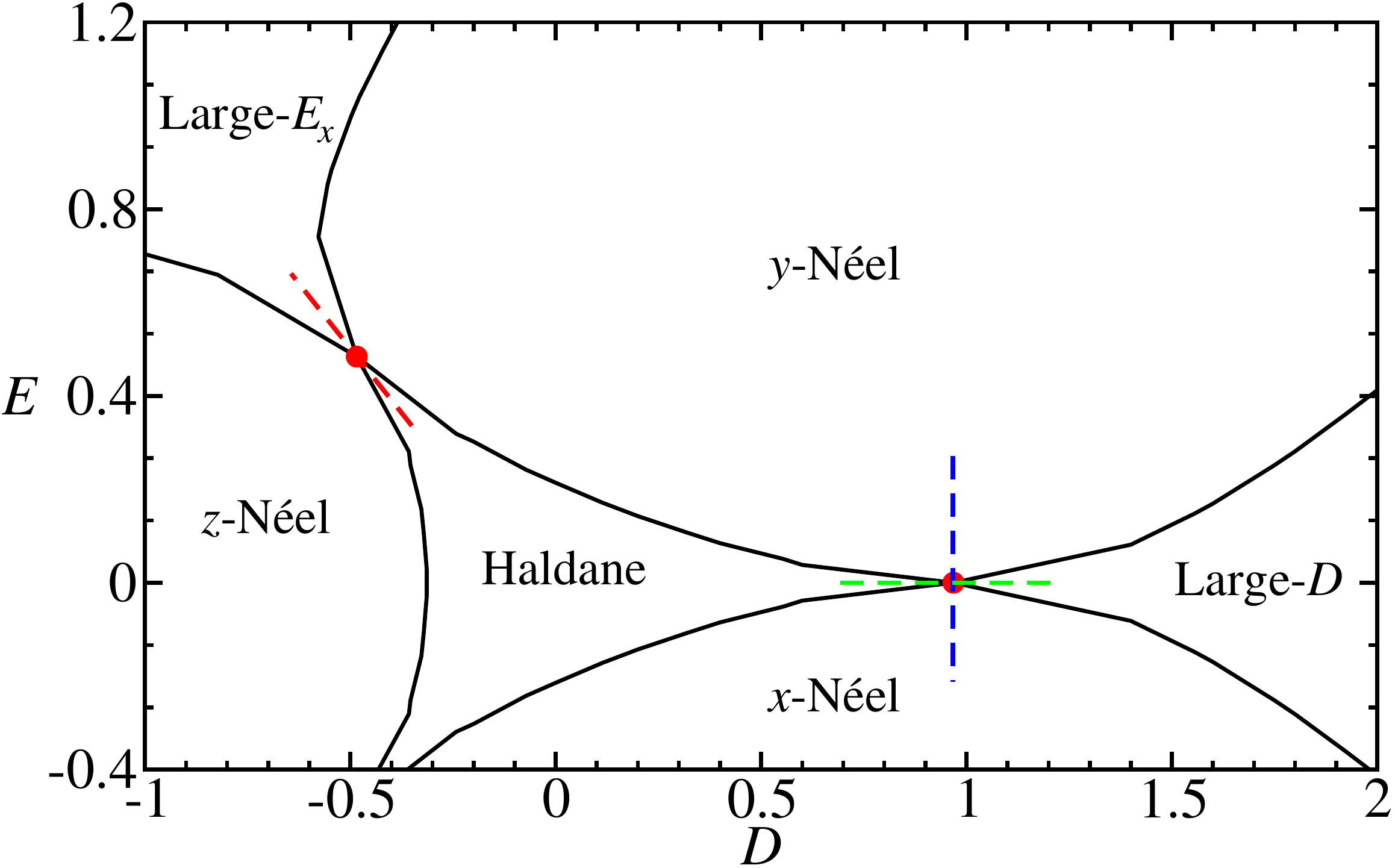}
      \caption{\label{Jz1PhaseDiagram}(Color online) The phase diagram of this model with $J_z=1$, where our main focus is labelled in different colors.}
    \end{center}
  \end{figure}

In the following two subsections, we will study the phase transitions in this model with $J_z=1$ by using the above cross derivative. A rough phase diagram is shown in Fig.\ \ref{Jz1PhaseDiagram} below to stress the focus, as labelled in different colors.

\subsection{$J_z=1$, 2nd-order phase transition}

  \begin{figure}[htbp]
    \begin{center}
      \includegraphics[width=0.45\textwidth,clip,angle=0]{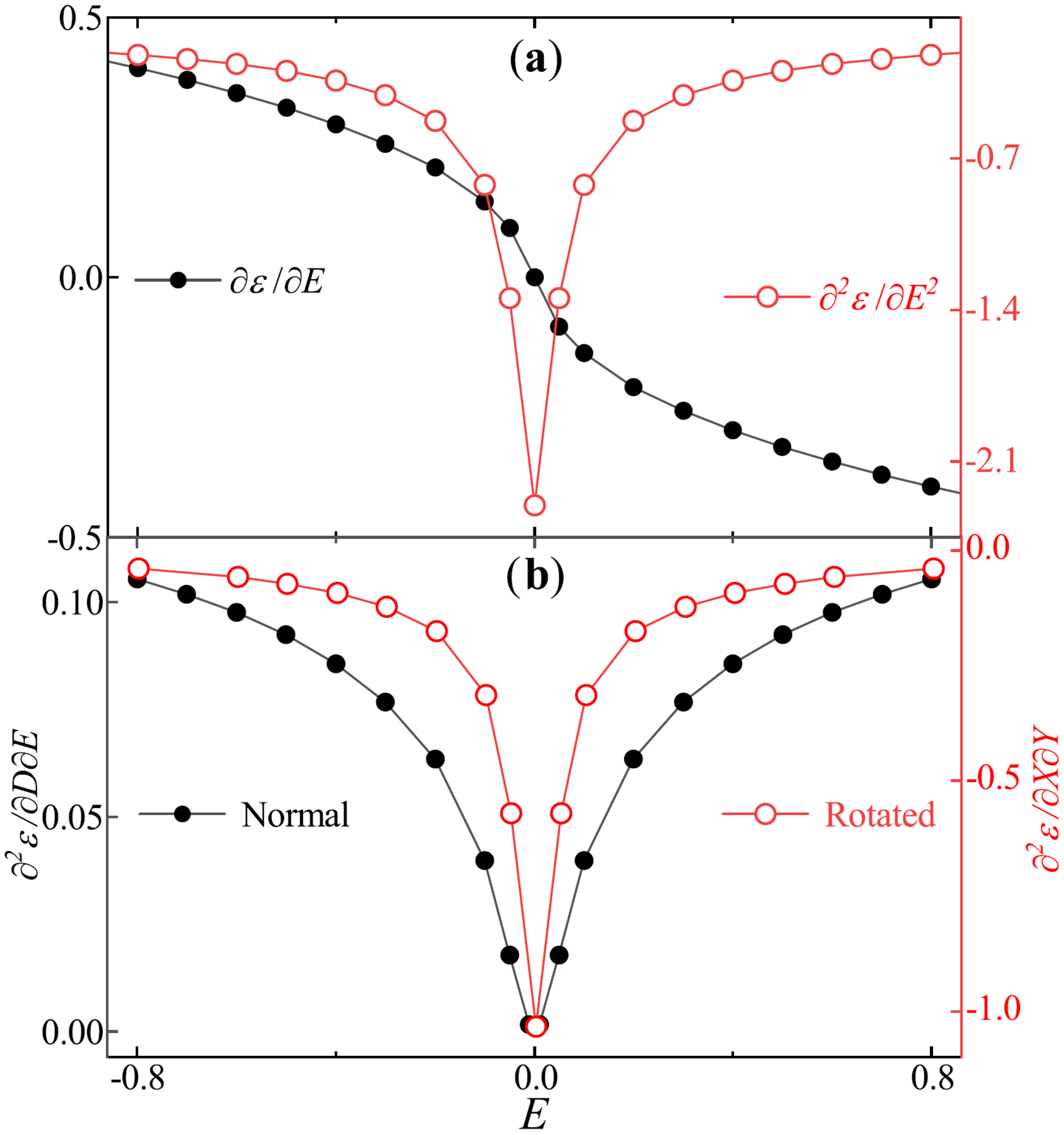}
      \caption{\label{compare}(Color online) For $L=40$ with fixing $D^\star=0.893730$: (a) 1st- and 2nd-order differentials of the ground state energy with varying $E$; (b) Both the normal and the rotated cross derivatives diverge at $E=0$.}
    \end{center}
  \end{figure}

When $J_z=1$ and $E=0$, the Eq. (\ref{eq1}) model reduces to the spin-1 Haldane chain with uniaxial anisotropy and has a 3rd-order Gaussian-type phase transition between the Haldane and the Large-$D$ phases at $D_{c1}\sim0.9684$\cite{ZengPRB2017, Wang2011}. As shown in Fig.\ \ref{Jz1PhaseDiagram}, in the $(D, E)$-plane, the $y$-N\'{e}el ($x$-N\'{e}el) phase lies
above (below) this point $(D_{c1}, 0)$, and between them, there is a 2nd-order quantum phase transition cross the point $(D_{c1}, 0)$ along the vertical $D=D_{c1}$ blue line.

Here, we first demonstrate the effectiveness of the cross derivative for this 2nd-order phase transition with the system size $L=40$. Fig.\ \ref{compare} shows the results, where the transition point $D^\star=0.893730$ is utilized (see Fig.\ \ref{MainResults1}(a)).
In Fig.\ \ref{compare}(a), one can clearly see that the 1st-order differential of the ground state energy $\partial{\epsilon}/{\partial{E}}$ is continuous, while the 2nd-order one $\partial^2{\epsilon}/{\partial{E^2}}$ is divergent at $E=0$, with $D$ fixed at $D^\star$. This is the smoking evidence of a 2nd-order phase transition.
In Fig.\ \ref{compare}(b), both the normal and the rotated cross derivatives show clear divergence at the same point $(D^\star, E=0)$, rendering identical information as in Fig.\ \ref{compare}(a). So, it shows clearly that in the study of 2nd-order phase transitions, the cross derivative can work equally well with the conventional derivative method.

\subsection{$J_z=1$, 3rd-order Gaussian-type transitions}

In this subsection, we will focus on the 3rd-order Gaussian-type quantum phase transitions in this model with $J_z=1$ and further illustrate the versatility of the proposed novel function.

  \begin{figure}[htbp]
    \begin{center}
      \includegraphics[width=0.42\textwidth,clip,angle=0]{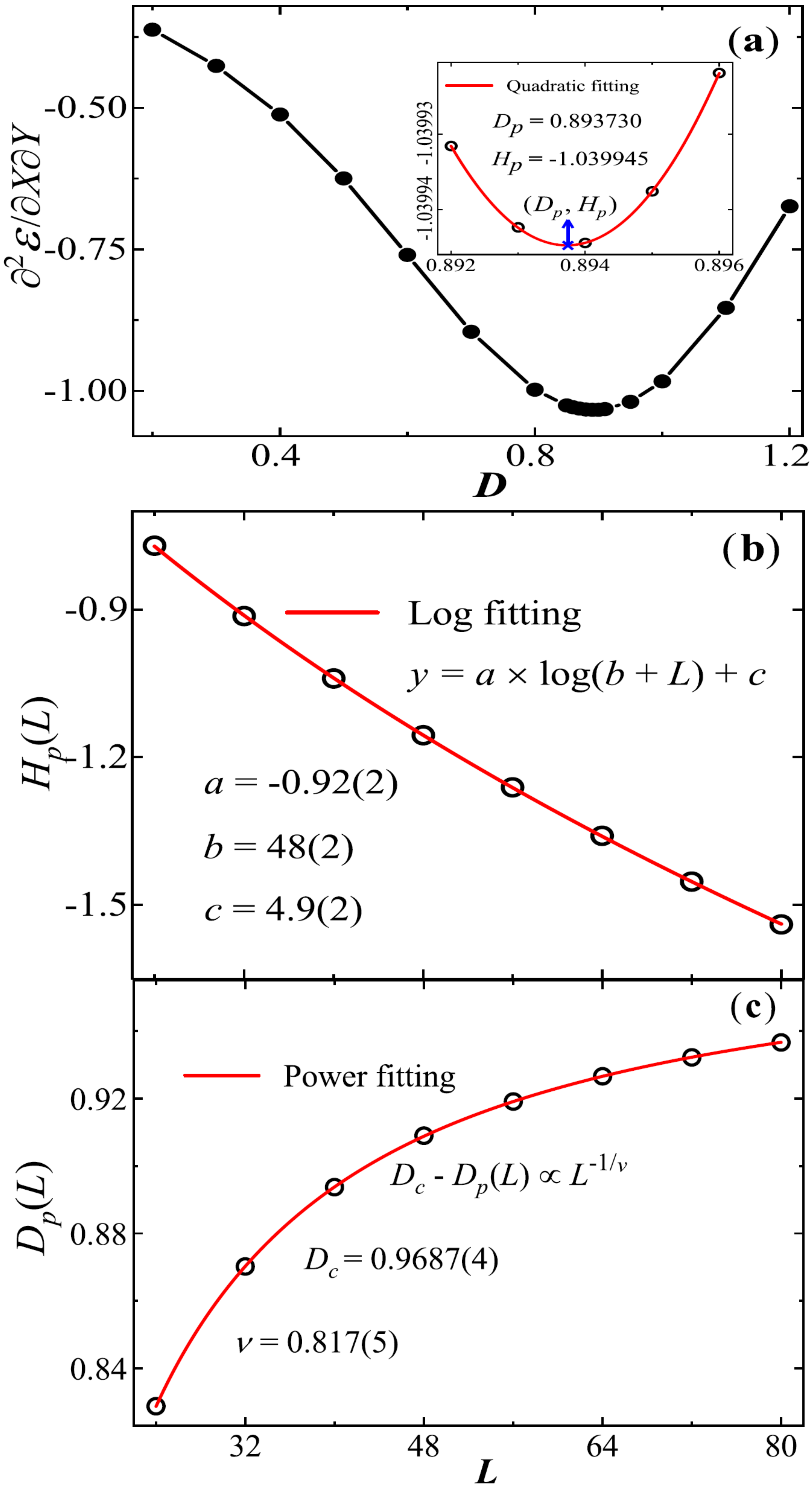}
      \caption{\label{MainResults1}(Color online) 
Phase transition between the Haldane phase and the Large-$D$ phase: (a) the rotated cross derivative for $L=40$ along the line $E=0$. The inset is a quadratic fitting near the minimum to obtain the valley position at $D_p=0.893730$ with depth $H_p=-1.039945$. (b) the valley depth $H_p(L)$ for different sizes with a logarithmic fitting. (c) the valley position $D_p(L)$ for different sizes with a power-law extrapolation. The estimated critical point locates at $D_{c1}=0.9687(4)$, and the critical exponent for the correlation length is $\nu=0.817(5)$.}
    \end{center}
  \end{figure}

As is well known, the 2nd-order differential of the ground state energy with respect to $D$, namely $\partial^2{\epsilon}/{\partial{D^2}}$, is incapable of detecting the transition between the Haldane and Large-$D$ phases, as shown in Fig.\ \ref{SD} of Sec.\ \ref{Appendix}, and in Fig.\ 3(c) of Ref.\ \onlinecite{Wang2011} and Fig.\ 4(b) of Ref.\ \onlinecite{ZengPRA22008}. More explicitly, Fig.\ \ref{SD}(b) has no power-law fitting property as the cross derivative does. Fig.\ 3(c) of Ref.\ \onlinecite{Wang2011} shows $\partial^2{\epsilon}/{\partial{D^2}}$ near the critical point, but neither any peak structure nor discontinuity shows up.
In Fig.\ 4(b) of Ref.\ \onlinecite{ZengPRA22008}, although $d^2\epsilon(L)/dD^2$ exhibits a broad peak-like structure, its location is somehow invariant as $L$ grows and is far away from the expected critical point. The peak height even decreases slightly instead of diverging, increasing $L$ from 100 to 220. 
None of these phenomena matches with a phase transition scenario.

According to Refs.\ [\onlinecite{ZengPRB2017,RenPRA2018,ZengPRA22008,Wang2011}], when $J_z=1$, there are indeed three 3rd-order phase transitions from the Haldane phase to the Large-$D$ phase at $(D_{c1}, 0)$, and two Large-$E$ phases at $(D_{c2}, \mp{D_{c2}})$ respectively, as shown in Fig.\ \ref{Jz1PhaseDiagram}. This also explains why the 2nd-order differential $\partial^2{\epsilon}/{\partial{D^2}}$ is inadequate to identify these transitions.
Here, we calculate the cross derivative ${\partial^2{\epsilon}}/{\partial{D}}{\partial{E}}$ (or ${\partial^2{\epsilon}}/{\partial{X}}{\partial{Y}}$) to investigate these transitions.
Since the Hamiltonian is invariant for opposite $E$s, the two Large-$E$ phases are symmetric about the line $E=0$, as well as the two Haldane-Large-$E$ critical points. So, we pick up one of these two.

  \begin{figure}[htbp]
    \begin{center}
      \includegraphics[width=0.42\textwidth,clip,angle=0]{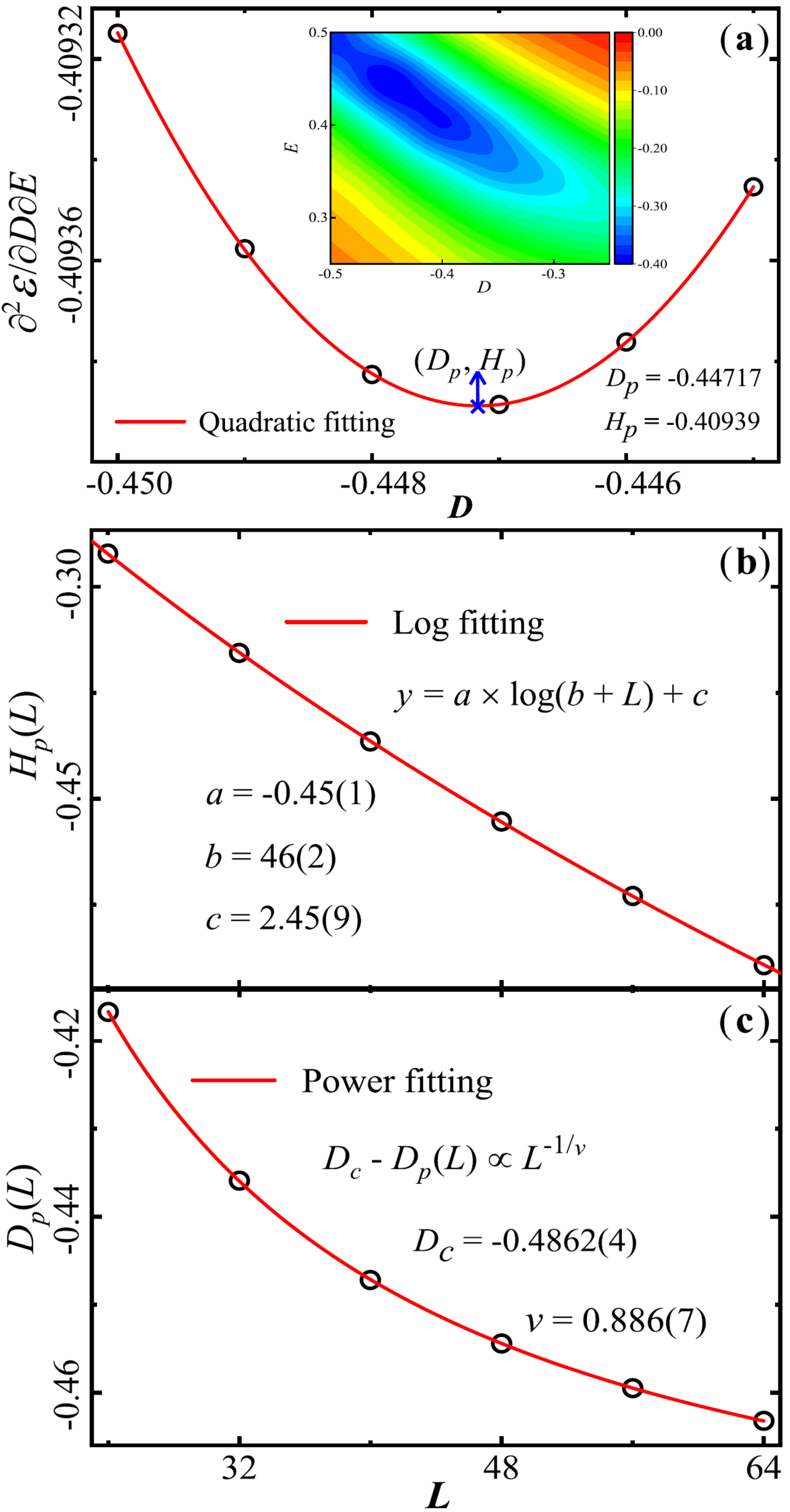}
      \caption{\label{MainResults2}(Color online) Phase transition between the Haldane phase and the Large-$E$ phase: (a) The normal cross derivative for $L=40$, and a quadratic fitting near the valley along the $E=-D$ direction is performed to locate the valley position at $D_p=-0.44717$ with depth $H_p=-0.40939$. The inset samples the $(D, E)$-plane. (b) The valley depth $H_p(L)$ for different sizes with a logarithmic fitting. (c) The valley position $D_p(L)$ for different sizes with a power-law fitting. The critical point is estimated at $D_{c2}=-0.4862(4)$, and the critical exponent for the correlation length is $\nu=0.886(7)$. }
    \end{center}
  \end{figure}

Figure\ \ref{MainResults1} shows the transition between the Haldane phase and the Large-$D$ phase on the line $E=0$, where the rotated cross derivative is adopted, as explained earlier. The result of $\partial^2\epsilon/\partial{X}\partial{Y}$ for $L=40$ is presented as an illustration in Fig.\ \ref{MainResults1}(a), where a single valley shows up clearly. In the inset, a quadratic fitting is performed around the valley minimum as,
\begin{equation}
  \partial^2\epsilon/\partial{X}\partial{Y} = a ( D - D_p(L) )^2 + H_p(L),
\end{equation}
and the valley depth is obtained as $H_p(L=40)=-1.039945$, with the location at $D_p(L=40)=0.893730$. We repeat this process for different system sizes from 24 to 80, and the valley depths for each size $H_p(L)$ are drawn in Fig.\ \ref{MainResults1}(b), which matches a logarithmic fitting perfectly as
\begin{equation}
  H_p(L) = a\ln(b+L)+c,
\end{equation}
with $a=-0.92(2)$, $b=48(2)$, and $c=4.9(2)$.
The logarithmic divergence of the valley depth with increasing the system size signals a phase transition therein. It then verifies the validity of the cross derivative for a phase transition.
Furthermore, the position of the valley minimum for each size $D_p(L)$ is collected in Fig.\ \ref{MainResults1}(c). Then, a power-law fitting\cite{FisherPRL} is performed as
\begin{equation}
   D_c - D_p(L) \propto L^{-1/\nu}.
\end{equation}
By the finite-size extrapolation, the critical point is determined at $D_{c1}=0.9687(4)$, and the critical exponent for the correlation length is $\nu=0.817(5)$. 
The estimated critical point is consistent with the most precise predictions to date\cite{ZengPRB2017,Wang2011,LyraPRB}, as listed in Tab.\ \ref{tableI} below. The critical exponent is smaller than $\nu=1.472$ from Ref.\  [\onlinecite{Wang2011}] and $\nu=1.42$ from Ref.\ [\onlinecite{ZengPRA22008}].

\begin{table}[!h]
  \begin{center}
    \caption{Comparison of $D_c$ estimated by different methods.}\label{tableI}
    \begin{tabular}{l c c}
    \hline\hline
      $D_c$  &   &  Method \\
          \hline
      0.95 &  & multi-target DMRG\cite{OrtolaniEPJB}\\          
      0.97 &  & fidelity(DMRG)\cite{ZengPRA2008} \\
      0.971(5) &  & stiffness(QMC)\cite{albuqPRB}  \\
      0.96845(8) &  & entropy(DMRG)\cite{Wang2011}\\
      0.9684713(1) &  & level spectroscopy(DMRG)\cite{ZengPRB2017}\\
      0.9685(2) &  & tangential finite size scaling\cite{LyraPRB}\\
      0.9687(4) &  & cross derivative(this work) \\
    \hline\hline
  \end{tabular}      
  \end{center}
\end{table}

For the Haldane-Large-$E$ transition, we first calculate the cross derivative by sampling in the $(D, E)$-plane, as shown in the inset of Fig.\ \ref{MainResults2}(a) for $L=40$.
Then, the transition point is estimated on the $E=-D$ line, as indicated by the inset contour, with a minimum around $D=-0.45=-E$. Finally, a quadratic fitting near the minimum is carried out to locate the valley position at $D_p=-0.44717$ with depth $H_p=-0.40939$. 
By collecting the valley depth $H_p(L)$ for different sizes from 24 to 64, we can observe a clear logarithmic divergence, as fitted in Fig.\ \ref{MainResults2}(b). Again, this logarithmic divergence of the valley depth with the increase of the system size indicates a phase transition. 
At the same time, with the valley position of each size $D_p(L)$ shown in Fig.\ \ref{MainResults2}(c), we obtain the critical point at $(D_{c2}, -D_{c2})$ with $D_{c2}=-0.4862(4)$ and the critical exponent for the correlation length as $\nu=0.886(7)$, by the finite size extrapolation. As expected, the transition point agrees well with the previous prediction\cite{ZengPRB2017}.


An interesting and worth mentioning thing is that the logarithmic fitting function here is precisely half of that in Fig.\ \ref{MainResults1}(b). The difference is consistent with the definitions, as sketched in Fig.\ \ref{CD_definition}. This fact that these two logarithmic fitting functions coincide well with each other verifies the two separate calculations are consistent and indicates the two transitions are the same type.

So far, we have shown that the proposed cross derivative of the Gibbs free energy can detect and locate the 3rd-order quantum Gaussian-type transitions in this model. Furthermore, convenience, precision, and efficiency are explicitly displayed.
Combining with Figs.\ \ref{compare} and \ref{SD}, we should note that the cross derivative contains contributions from both orthogonal directions, where the divergence behavior comes from the nature of the 2nd-order transition in $E$-direction, as expressed in Eq. (\ref{eq4}). This may explain the reason why the cross derivative works well.

  \begin{figure}[htbp]
    \begin{center}
      \includegraphics[width=0.42\textwidth,clip,angle=0]{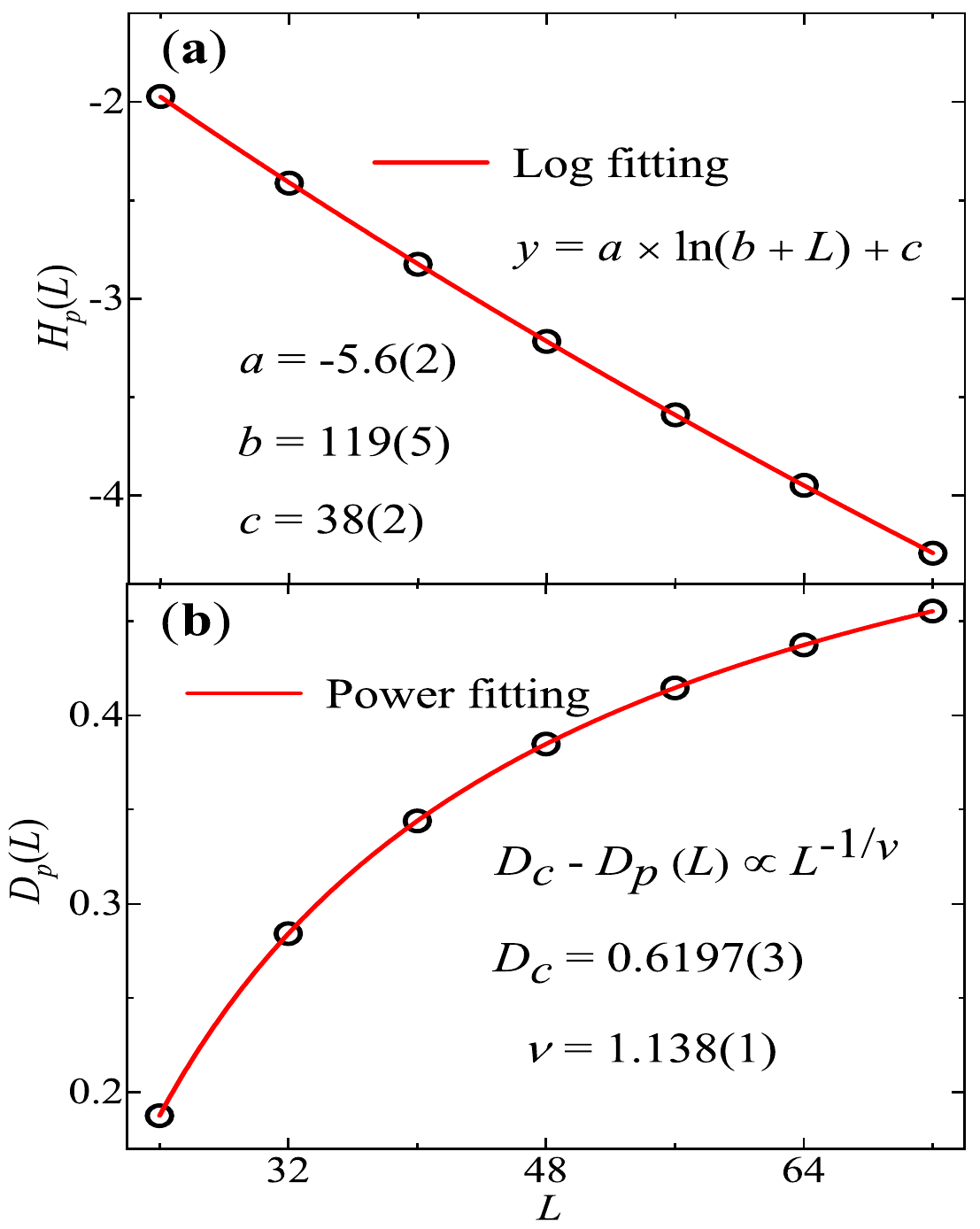}
      \caption{\label{MainResults3}(Color online) When $J_z=0.5$, the  rotated cross derivative is computed along the line $E=0$: (a) the valley depth $H_p(L)$ for different sizes with a perfect logarithmic fitting; (b) the location of the valley minimum $D_p(L)$ with a power-law fitting. The critical point locates at $D_{c}=0.6197(3)$, and the critical exponent for the correlation length is $\nu = 1.138(1) $.}
    \end{center}
  \end{figure}

\subsection{$J_z=0.5$, 5th-order Gaussian-type transition}

In this model with $J_z=0.5$, a 5th-order Gaussian-type quantum phase transition between the Haldane phase and the Large-$D$ phase has also been reported along $E=0$ at $D_c\approx0.63$\cite{ZengPRA22008, BoschiEPJB2003,Wang2011,ChenPRB2003}. However, this transition and the critical point are even more challenging to detect and locate. Recent researches have adopted the entanglement entropy, the fidelity susceptibility, or other complex quantities.

 We follow the same logic for this case, and compute the rotated cross derivative ${\partial^2{\epsilon}}/{\partial{X}}{\partial{Y}}$ by fixing $E=0$. The results are presented in Fig.\ \ref{MainResults3}, and the valley lies at $D_p=0.3500$ with depth $H_p=-2.862$ for $L=40$. 
A logarithmic fitting is performed from the collected data of $H_p(L)$ with $L$ ranging from 24 to 72. They match each other perfectly, and the logarithmic divergence denotes the phase transition. By collecting $D_p(L)$ with different sizes, a finite size extrapolation is carried out to obtain the critical point at $D_c=0.6197(3)$ and the critical exponent for the correlation length as $\nu=1.138(1)$ simultaneously. Both are pretty close to the estimations of $D_c=0.63$ and $\nu=1.51$ in Ref.\ [\onlinecite{ZengPRA22008}], and $D_c=0.635$ in Refs.\ [\onlinecite{Wang2011,ChenPRB2003}]. 

Once again, the validity and efficiency of the cross derivative are illustrated for this difficult 5th-order Gaussian-type quantum phase transition.

\section{Summary and discussion}\label{Summary}

In brief, we extend the scope of the cross derivative of the Gibbs free energy, initially proposed for phase transitions in classical spin models\cite{YuPRB2020}, to the study of quantum cases. Its validity and efficiency have been demonstrated by the typical and challenging higher-order Gaussian-type phase transitions in spin-1 {\textit{XXZ}} chain with anisotropies.

When $J_z=1$, the 3rd-order Gaussian-type phase transition is precisely located at (0.9687, 0) for the Haldane-Large-$D$ transition and at (-0.4862, 0.4862) between the Haldane phase and the Large-$E$ phase. The obtained critical points agree well with the most accurate estimations to date in the literature. As for $J_z=0.5$, the 5th-order Gaussian-type transition is determined at $(0.6194, 0)$, also consistent with the previous predictions.

In both transitions, the critical exponent for the correlation length is a little smaller than the predictions in the literature. We should note that, for higher-order continuous phase transition, the critical exponent for the correlation length is more difficult to determine precisely than the location of the critical point. When approaching the critical point, the correlation length grows rapidly and becomes much larger than the size used in this work. To accurately estimate the critical exponent for the correlation length, one may need larger system sizes for the finite size scaling to eliminate the small size effect, as mentioned in Ref.\ [\onlinecite{Wang2011}], wherein the biggest system size utilized is $10^4$. To further improve the accuracy of the estimated critical exponent with a larger system size, one may try other numerical algorithms, like the recently developed variational corner transfer matrix renormalization group method \cite{vCTMRG} or the (infinite) time-evolving block decimation method \cite{vidal_TEBDPRL2004,vidal_iTEBDPRL2007,vidal_iTEBDPRB2008} to even deal with an infinite one-dimensional system directly. The other minor possibility is the equal weight (i.e., $\alpha=\beta$) adopted in Eq. (\ref{eq2}) for simplicity, which may not be optimal and requires further investigations. These will be presented in subsequent studies. Also, 
the order of the quantum critical point can be altered by changing the direction in the phase diagram across the critical point, which may also affect the critical exponent.

Given its simplicity and convenience, the cross derivative is efficient and universal to investigate the phase transitions in quantum spin systems, whether it is the conventional 2nd-order one or the complicated Gaussian-type one. Moreover, the predictions will be more accurate if the system size and the precision of the ground state energy can be improved further. 
The method is also readily applied to other complex systems, like the argued 3rd-order phase transition in Bose-Einstein condensation\cite{BEC1,BEC2, BEC3}.
Requiring only the precise ground state energy, this quantity is much easier to deal with than wave function, correlation functions or order parameters used in other methods.

\section{Acknowledgement}

Y.-C.T. acknowledges the support from National Center for Theoretical Sciences (NCTS) in Taiwan.
Z.Y.X. is supported by National R\&D Program of China (Grants Nos. 2017YFA0302900, 2016YFA0300503), National Natural Science Foundation of China (Grants Nos. 12274458, 11774420), and the Research Funds of Renmin University of China (Grant No. 20XNLG19). 
K.J. is supported by the National Natural Science Foundation of China (Grant No. 11974249).
J.F.Y. is supported by Natural Science Foundation of Hunan Province (No. 851204035).

H.Y.W. and Y.-C.T. contributed equally to this work. 



\section{APPENDIX: 2nd-order differential of $\epsilon$ along $D$-direction}\label{Appendix}

\begin{figure}[htbp]
    \centering
    \subfigure{\includegraphics[width=0.44\textwidth,clip,angle=0]{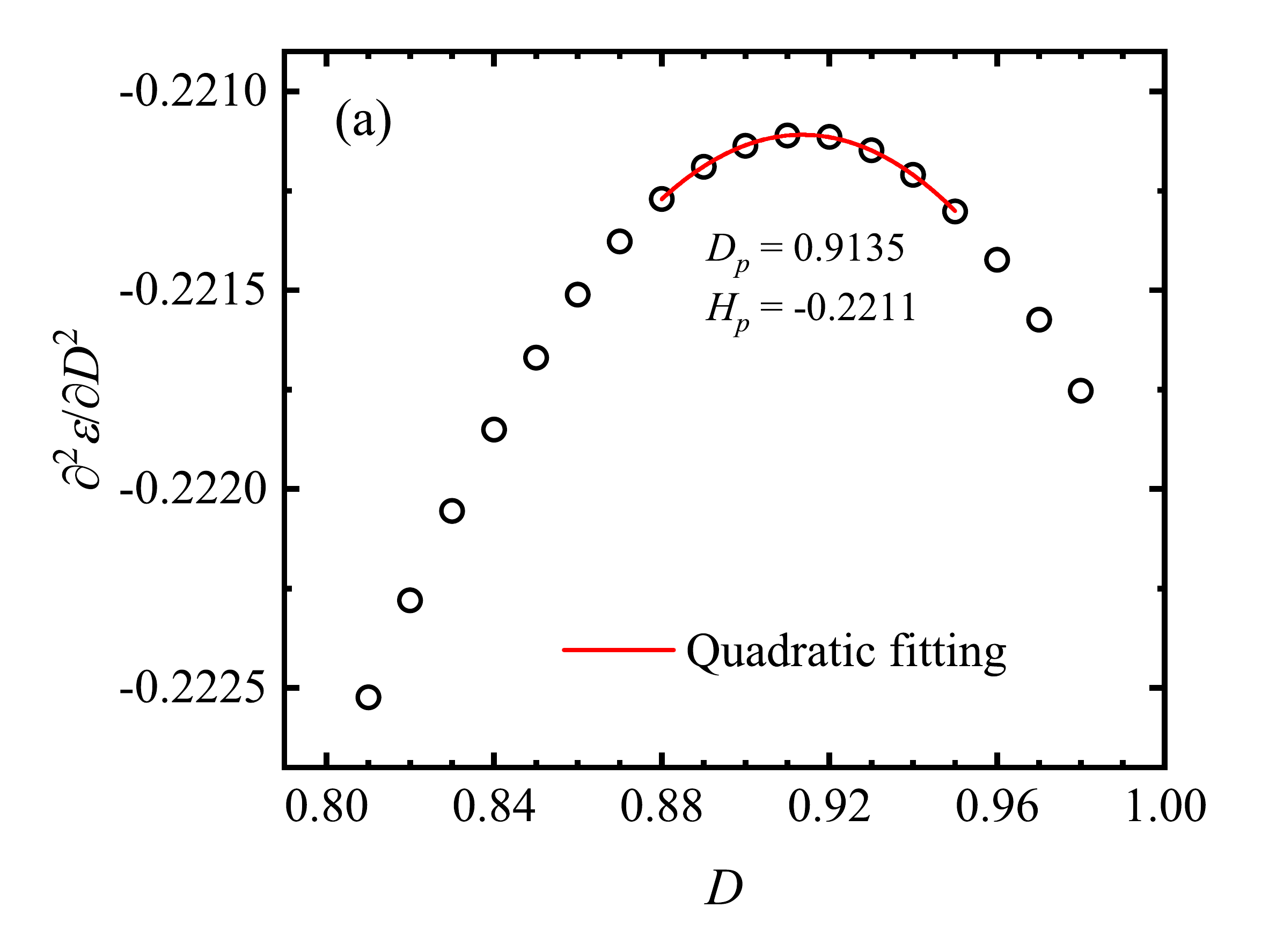}\label{xxx1}}
    \vspace{0.5in}
\subfigure{\includegraphics[width=0.42\textwidth,clip,angle=0]{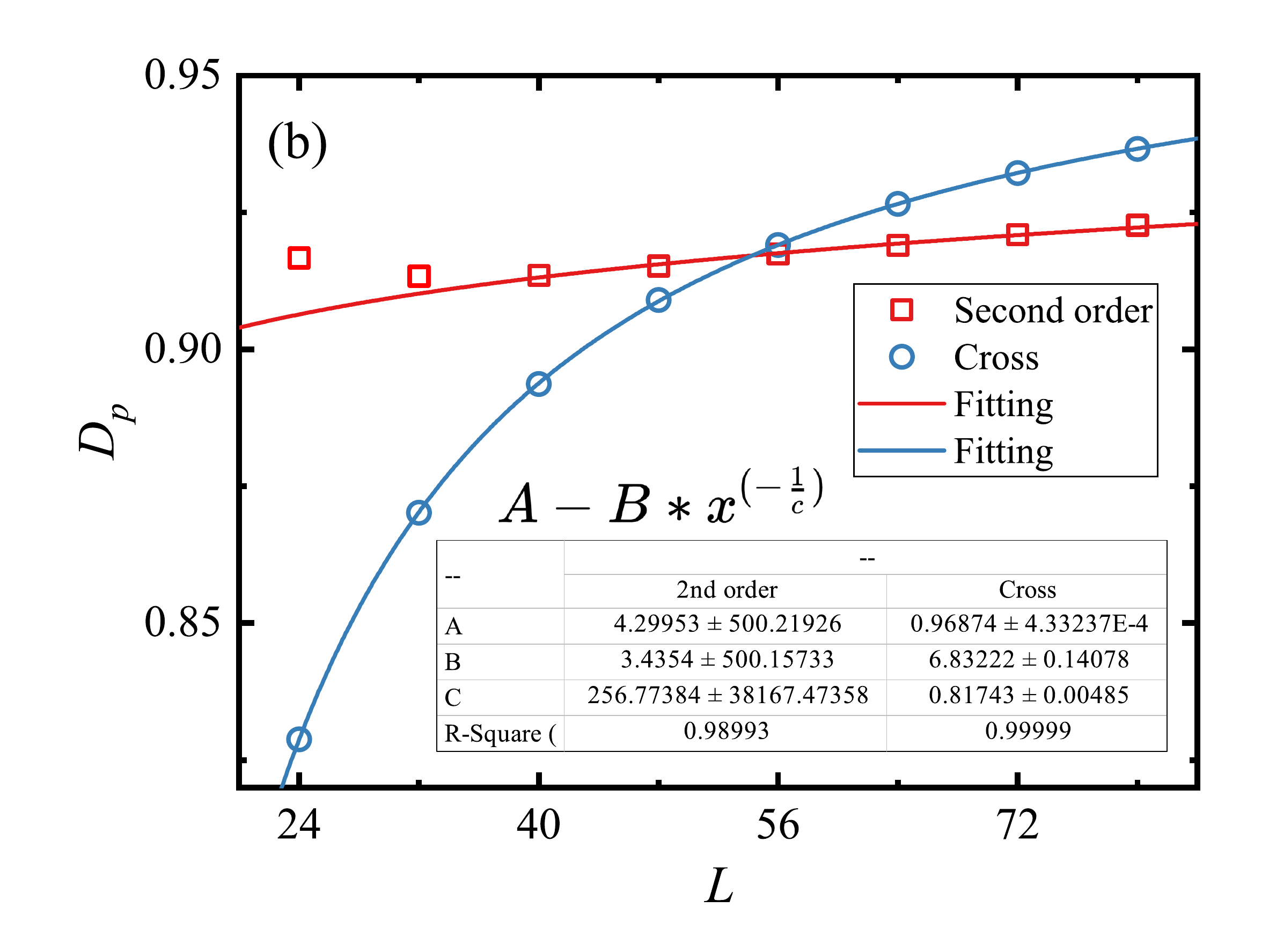}\label{xxx2}}
    \caption{\label{SD}(a) Illustration of the 2nd order differential $\partial^2\epsilon/\partial{D^2}$along $D$ direction with $L=40$, and a quadratic fitting near the peak to give the peak position; (b) Power fitting of the peak position of the differential $\partial^2\epsilon/\partial{D^2}$, and the fitting still fails  even we neglect two values for small size systems. For comparison, the rotated cross derivative, i.e. Fig.\ \ref{MainResults1}(c) is also included in blue.}
\end{figure}

According to the statistical physics, an $n$-th order phase transition is detected from the divergence of the $n$-th order differential of the Gibbs free energy, while whose $(n-1)$-th order differential is continuous. So naturally, a 2nd-order differential is not able to identify a higher-order phase transition. Here, we show below in Fig.\ \ref{SD} the failure of the 2nd order differential of the Gibbs free energy with respect to $D$, to detect the 3rd order Gaussian transition in this model along $D$ direction when $J_z=1$. For comparison, the rotated cross derivative (Fig.\ \ref{MainResults1}(c)) is also included.

\bibliography{XXZDE}

\providecommand{\noopsort}[1]{}\providecommand{\singleletter}[1]{#1}%
\begin{thebibliography}{34}%
\makeatletter
\providecommand \@ifxundefined [1]{%
 \@ifx{#1\undefined}
}%
\providecommand \@ifnum [1]{%
 \ifnum #1\expandafter \@firstoftwo
 \else \expandafter \@secondoftwo
 \fi
}%
\providecommand \@ifx [1]{%
 \ifx #1\expandafter \@firstoftwo
 \else \expandafter \@secondoftwo
 \fi
}%
\providecommand \natexlab [1]{#1}%
\providecommand \enquote  [1]{``#1''}%
\providecommand \bibnamefont  [1]{#1}%
\providecommand \bibfnamefont [1]{#1}%
\providecommand \citenamefont [1]{#1}%
\providecommand \href@noop [0]{\@secondoftwo}%
\providecommand \href [0]{\begingroup \@sanitize@url \@href}%
\providecommand \@href[1]{\@@startlink{#1}\@@href}%
\providecommand \@@href[1]{\endgroup#1\@@endlink}%
\providecommand \@sanitize@url [0]{\catcode `\\12\catcode `\$12\catcode
  `\&12\catcode `\#12\catcode `\^12\catcode `\_12\catcode `\%12\relax}%
\providecommand \@@startlink[1]{}%
\providecommand \@@endlink[0]{}%
\providecommand \url  [0]{\begingroup\@sanitize@url \@url }%
\providecommand \@url [1]{\endgroup\@href {#1}{\urlprefix }}%
\providecommand \urlprefix  [0]{URL }%
\providecommand \Eprint [0]{\href }%
\providecommand \doibase [0]{http://dx.doi.org/}%
\providecommand \selectlanguage [0]{\@gobble}%
\providecommand \bibinfo  [0]{\@secondoftwo}%
\providecommand \bibfield  [0]{\@secondoftwo}%
\providecommand \translation [1]{[#1]}%
\providecommand \BibitemOpen [0]{}%
\providecommand \bibitemStop [0]{}%
\providecommand \bibitemNoStop [0]{.\EOS\space}%
\providecommand \EOS [0]{\spacefactor3000\relax}%
\providecommand \BibitemShut  [1]{\csname bibitem#1\endcsname}%
\let\auto@bib@innerbib\@empty
\bibitem [{\citenamefont {Kosterlitz}\ and\ \citenamefont
  {Thouless}(1973)}]{KT1}%
  \BibitemOpen
  \bibfield  {author} {\bibinfo {author} {\bibfnamefont {J.~M.}\ \bibnamefont
  {Kosterlitz}}\ and\ \bibinfo {author} {\bibfnamefont {D.~J.}\ \bibnamefont
  {Thouless}},\ }\href {\doibase 10.1088/0022-3719/6/7/010} {\bibfield
  {journal} {\bibinfo  {journal} {J. Phys. C}\ }\textbf {\bibinfo {volume}
  {6}},\ \bibinfo {pages} {1181} (\bibinfo {year} {1973})}\BibitemShut
  {NoStop}%
\bibitem [{\citenamefont {Kosterlitz}(1974)}]{KT2}%
  \BibitemOpen
  \bibfield  {author} {\bibinfo {author} {\bibfnamefont {J.~M.}\ \bibnamefont
  {Kosterlitz}},\ }\href {\doibase 10.1088/0022-3719/7/6/005} {\bibfield
  {journal} {\bibinfo  {journal} {J. Phys. C}\ }\textbf {\bibinfo {volume}
  {7}},\ \bibinfo {pages} {1046} (\bibinfo {year} {1974})}\BibitemShut
  {NoStop}%
\bibitem [{\citenamefont {Haldane}(1983{\natexlab{a}})}]{Haldane1}%
  \BibitemOpen
  \bibfield  {author} {\bibinfo {author} {\bibfnamefont {F.~D.~M.}\
  \bibnamefont {Haldane}},\ }\href
  {https://doi.org/10.1016/0375-9601(83)90631-X} {\bibfield  {journal}
  {\bibinfo  {journal} {Phys. Lett. A}\ }\textbf {\bibinfo {volume} {93}}
  (\bibinfo {year} {1983}{\natexlab{a}})}\BibitemShut {NoStop}%
\bibitem [{\citenamefont {Haldane}(1983{\natexlab{b}})}]{Haldane2}%
  \BibitemOpen
  \bibfield  {author} {\bibinfo {author} {\bibfnamefont {F.~D.~M.}\
  \bibnamefont {Haldane}},\ }\href
  {chttps://doi.org/10.1103/PhysRevLett.50.1153} {\bibfield  {journal}
  {\bibinfo  {journal} {Phys. Rev. Lett.}\ }\textbf {\bibinfo {volume} {50}}
  (\bibinfo {year} {1983}{\natexlab{b}})}\BibitemShut {NoStop}%
\bibitem [{\citenamefont {Haldane}(2017)}]{Haldane3}%
  \BibitemOpen
  \bibfield  {author} {\bibinfo {author} {\bibfnamefont {F.~D.~M.}\
  \bibnamefont {Haldane}},\ }\href
  {https://doi.org/10.1016/0375-9601(83)90631-X} {\bibfield  {journal}
  {\bibinfo  {journal} {Rev. Mod. Phys.}\ }\textbf {\bibinfo {volume} {89}}
  (\bibinfo {year} {2017})}\BibitemShut {NoStop}%
\bibitem [{\citenamefont {Chen}\ \emph {et~al.}(2020)\citenamefont {Chen},
  \citenamefont {Ji}, \citenamefont {Xie},\ and\ \citenamefont
  {Yu}}]{YuPRB2020}%
  \BibitemOpen
  \bibfield  {author} {\bibinfo {author} {\bibfnamefont {Y.}~\bibnamefont
  {Chen}}, \bibinfo {author} {\bibfnamefont {K.}~\bibnamefont {Ji}}, \bibinfo
  {author} {\bibfnamefont {Z.~Y.}\ \bibnamefont {Xie}}, \ and\ \bibinfo
  {author} {\bibfnamefont {J.~F.}\ \bibnamefont {Yu}},\ }\href {\doibase
  10.1103/PhysRevB.101.165123} {\bibfield  {journal} {\bibinfo  {journal}
  {Phys. Rev. B}\ }\textbf {\bibinfo {volume} {101}},\ \bibinfo {pages}
  {165123} (\bibinfo {year} {2020})}\BibitemShut {NoStop}%
\bibitem [{\citenamefont {Tzeng}\ \emph {et~al.}(2017)\citenamefont {Tzeng},
  \citenamefont {Onishi}, \citenamefont {Okubo},\ and\ \citenamefont
  {Kao}}]{ZengPRB2017}%
  \BibitemOpen
  \bibfield  {author} {\bibinfo {author} {\bibfnamefont {Y.-C.}\ \bibnamefont
  {Tzeng}}, \bibinfo {author} {\bibfnamefont {H.}~\bibnamefont {Onishi}},
  \bibinfo {author} {\bibfnamefont {T.}~\bibnamefont {Okubo}}, \ and\ \bibinfo
  {author} {\bibfnamefont {Y.-J.}\ \bibnamefont {Kao}},\ }\href {\doibase
  10.1103/PhysRevB.96.060404} {\bibfield  {journal} {\bibinfo  {journal} {Phys.
  Rev. B}\ }\textbf {\bibinfo {volume} {96}},\ \bibinfo {pages} {060404(R)}
  (\bibinfo {year} {2017})}\BibitemShut {NoStop}%
\bibitem [{\citenamefont {Hu}\ \emph {et~al.}(2011)\citenamefont {Hu},
  \citenamefont {Normand}, \citenamefont {Wang},\ and\ \citenamefont
  {Yu}}]{Wang2011}%
  \BibitemOpen
  \bibfield  {author} {\bibinfo {author} {\bibfnamefont {S.}~\bibnamefont
  {Hu}}, \bibinfo {author} {\bibfnamefont {B.}~\bibnamefont {Normand}},
  \bibinfo {author} {\bibfnamefont {X.}~\bibnamefont {Wang}}, \ and\ \bibinfo
  {author} {\bibfnamefont {L.}~\bibnamefont {Yu}},\ }\href {\doibase
  10.1103/PhysRevB.84.220402} {\bibfield  {journal} {\bibinfo  {journal} {Phys.
  Rev. B}\ }\textbf {\bibinfo {volume} {84}},\ \bibinfo {pages} {220402}
  (\bibinfo {year} {2011})}\BibitemShut {NoStop}%
\bibitem [{\citenamefont {Tzeng}\ \emph {et~al.}(2008)\citenamefont {Tzeng},
  \citenamefont {Hung}, \citenamefont {Chen},\ and\ \citenamefont
  {Yang}}]{ZengPRA22008}%
  \BibitemOpen
  \bibfield  {author} {\bibinfo {author} {\bibfnamefont {Y.-C.}\ \bibnamefont
  {Tzeng}}, \bibinfo {author} {\bibfnamefont {H.-H.}\ \bibnamefont {Hung}},
  \bibinfo {author} {\bibfnamefont {Y.-C.}\ \bibnamefont {Chen}}, \ and\
  \bibinfo {author} {\bibfnamefont {M.-F.}\ \bibnamefont {Yang}},\ }\href
  {\doibase 10.1103/PhysRevA.77.062321} {\bibfield  {journal} {\bibinfo
  {journal} {Phys. Rev. A}\ }\textbf {\bibinfo {volume} {77}},\ \bibinfo
  {pages} {062321} (\bibinfo {year} {2008})}\BibitemShut {NoStop}%
\bibitem [{\citenamefont {{C. Degli Esposti Boschi}}\ \emph
  {et~al.}(2003)\citenamefont {{C. Degli Esposti Boschi}}, \citenamefont {{E.
  Ercolessi}}, \citenamefont {{F. Ortolani}},\ and\ \citenamefont {{M.
  Roncaglia}}}]{BoschiEPJB2003}%
  \BibitemOpen
  \bibfield  {author} {\bibinfo {author} {\bibnamefont {{C. Degli Esposti
  Boschi}}}, \bibinfo {author} {\bibnamefont {{E. Ercolessi}}}, \bibinfo
  {author} {\bibnamefont {{F. Ortolani}}}, \ and\ \bibinfo {author}
  {\bibnamefont {{M. Roncaglia}}},\ }\href {\doibase
  10.1140/epjb/e2003-00299-7} {\bibfield  {journal} {\bibinfo  {journal} {Eur.
  Phys. J. B}\ }\textbf {\bibinfo {volume} {35}},\ \bibinfo {pages} {465}
  (\bibinfo {year} {2003})}\BibitemShut {NoStop}%
\bibitem [{\citenamefont {Tzeng}\ and\ \citenamefont
  {Yang}(2008)}]{ZengPRA2008}%
  \BibitemOpen
  \bibfield  {author} {\bibinfo {author} {\bibfnamefont {Y.-C.}\ \bibnamefont
  {Tzeng}}\ and\ \bibinfo {author} {\bibfnamefont {M.-F.}\ \bibnamefont
  {Yang}},\ }\href {\doibase 10.1103/PhysRevA.77.012311} {\bibfield  {journal}
  {\bibinfo  {journal} {Phys. Rev. A}\ }\textbf {\bibinfo {volume} {77}},\
  \bibinfo {pages} {012311} (\bibinfo {year} {2008})}\BibitemShut {NoStop}%
\bibitem [{\citenamefont {Maximova}\ \emph {et~al.}(2021)\citenamefont
  {Maximova}, \citenamefont {Streltsov},\ and\ \citenamefont
  {Vasiliev}}]{Spin1:compounds}%
  \BibitemOpen
  \bibfield  {author} {\bibinfo {author} {\bibfnamefont {O.~V.}\ \bibnamefont
  {Maximova}}, \bibinfo {author} {\bibfnamefont {S.~V.}\ \bibnamefont
  {Streltsov}}, \ and\ \bibinfo {author} {\bibfnamefont {A.~N.}\ \bibnamefont
  {Vasiliev}},\ }\href {\doibase 10.1080/10408436.2020.1852911} {\bibfield
  {journal} {\bibinfo  {journal} {Crit. Rev. Solid State Mat. Sci.}\ }\textbf
  {\bibinfo {volume} {46}},\ \bibinfo {pages} {371} (\bibinfo {year}
  {2021})}\BibitemShut {NoStop}%
\bibitem [{\citenamefont {Chung}\ \emph {et~al.}(2021)\citenamefont {Chung},
  \citenamefont {de~Hond}, \citenamefont {Xiang}, \citenamefont {Cruz-Colon},\
  and\ \citenamefont {Ketterle}}]{Woo}%
  \BibitemOpen
  \bibfield  {author} {\bibinfo {author} {\bibfnamefont {W.~C.}\ \bibnamefont
  {Chung}}, \bibinfo {author} {\bibfnamefont {J.}~\bibnamefont {de~Hond}},
  \bibinfo {author} {\bibfnamefont {J.}~\bibnamefont {Xiang}}, \bibinfo
  {author} {\bibfnamefont {E.}~\bibnamefont {Cruz-Colon}}, \ and\ \bibinfo
  {author} {\bibfnamefont {W.}~\bibnamefont {Ketterle}},\ }\href {\doibase
  10.1103/PhysRevLett.126.163203} {\bibfield  {journal} {\bibinfo  {journal}
  {Phys. Rev. Lett.}\ }\textbf {\bibinfo {volume} {126}},\ \bibinfo {pages}
  {163203} (\bibinfo {year} {2021})}\BibitemShut {NoStop}%
\bibitem [{\citenamefont {Pajerowski}\ \emph {et~al.}(2022)\citenamefont
  {Pajerowski}, \citenamefont {Podlesnyak}, \citenamefont {Herbrych},\ and\
  \citenamefont {Manson}}]{NBCT}%
  \BibitemOpen
  \bibfield  {author} {\bibinfo {author} {\bibfnamefont {D.~M.}\ \bibnamefont
  {Pajerowski}}, \bibinfo {author} {\bibfnamefont {A.~P.}\ \bibnamefont
  {Podlesnyak}}, \bibinfo {author} {\bibfnamefont {J.}~\bibnamefont
  {Herbrych}}, \ and\ \bibinfo {author} {\bibfnamefont {J.}~\bibnamefont
  {Manson}},\ }\href {\doibase 10.1103/PhysRevB.105.134420} {\bibfield
  {journal} {\bibinfo  {journal} {Phys. Rev. B}\ }\textbf {\bibinfo {volume}
  {105}},\ \bibinfo {pages} {134420} (\bibinfo {year} {2022})}\BibitemShut
  {NoStop}%
\bibitem [{\citenamefont {Ren}\ \emph {et~al.}(2018)\citenamefont {Ren},
  \citenamefont {Wang},\ and\ \citenamefont {You}}]{RenPRA2018}%
  \BibitemOpen
  \bibfield  {author} {\bibinfo {author} {\bibfnamefont {J.}~\bibnamefont
  {Ren}}, \bibinfo {author} {\bibfnamefont {Y.}~\bibnamefont {Wang}}, \ and\
  \bibinfo {author} {\bibfnamefont {W.-L.}\ \bibnamefont {You}},\ }\href
  {\doibase 10.1103/PhysRevA.97.042318} {\bibfield  {journal} {\bibinfo
  {journal} {Phys. Rev. A}\ }\textbf {\bibinfo {volume} {97}},\ \bibinfo
  {pages} {042318} (\bibinfo {year} {2018})}\BibitemShut {NoStop}%
\bibitem [{\citenamefont {Johnston}(2017)}]{JohnstonPRB}%
  \BibitemOpen
  \bibfield  {author} {\bibinfo {author} {\bibfnamefont {D.~C.}\ \bibnamefont
  {Johnston}},\ }\href {\doibase 10.1103/PhysRevB.95.094421} {\bibfield
  {journal} {\bibinfo  {journal} {Phys. Rev. B}\ }\textbf {\bibinfo {volume}
  {95}},\ \bibinfo {pages} {094421} (\bibinfo {year} {2017})}\BibitemShut
  {NoStop}%
\bibitem [{\citenamefont {Albuquerque}\ \emph {et~al.}(2009)\citenamefont
  {Albuquerque}, \citenamefont {Hamer},\ and\ \citenamefont
  {Oitmaa}}]{albuqPRB}%
  \BibitemOpen
  \bibfield  {author} {\bibinfo {author} {\bibfnamefont {A.~F.}\ \bibnamefont
  {Albuquerque}}, \bibinfo {author} {\bibfnamefont {C.~J.}\ \bibnamefont
  {Hamer}}, \ and\ \bibinfo {author} {\bibfnamefont {J.}~\bibnamefont
  {Oitmaa}},\ }\href {\doibase 10.1103/PhysRevB.79.054412} {\bibfield
  {journal} {\bibinfo  {journal} {Phys. Rev. B}\ }\textbf {\bibinfo {volume}
  {79}},\ \bibinfo {pages} {054412} (\bibinfo {year} {2009})}\BibitemShut
  {NoStop}%
\bibitem [{\citenamefont {Liu}\ \emph {et~al.}(2014)\citenamefont {Liu},
  \citenamefont {Li}, \citenamefont {You}, \citenamefont {Su},\ and\
  \citenamefont {Tian}}]{GHLiuPB}%
  \BibitemOpen
  \bibfield  {author} {\bibinfo {author} {\bibfnamefont {G.-H.}\ \bibnamefont
  {Liu}}, \bibinfo {author} {\bibfnamefont {W.}~\bibnamefont {Li}}, \bibinfo
  {author} {\bibfnamefont {W.-L.}\ \bibnamefont {You}}, \bibinfo {author}
  {\bibfnamefont {G.}~\bibnamefont {Su}}, \ and\ \bibinfo {author}
  {\bibfnamefont {G.-S.}\ \bibnamefont {Tian}},\ }\href@noop {} {\bibfield
  {journal} {\bibinfo  {journal} {Physica B}\ }\textbf {\bibinfo {volume}
  {443}},\ \bibinfo {pages} {63} (\bibinfo {year} {2014})}\BibitemShut
  {NoStop}%
\bibitem [{\citenamefont {Chen}\ \emph {et~al.}(2003)\citenamefont {Chen},
  \citenamefont {Hida},\ and\ \citenamefont {Sanctuary}}]{ChenPRB2003}%
  \BibitemOpen
  \bibfield  {author} {\bibinfo {author} {\bibfnamefont {W.}~\bibnamefont
  {Chen}}, \bibinfo {author} {\bibfnamefont {K.}~\bibnamefont {Hida}}, \ and\
  \bibinfo {author} {\bibfnamefont {B.~C.}\ \bibnamefont {Sanctuary}},\ }\href
  {\doibase 10.1103/PhysRevB.67.104401} {\bibfield  {journal} {\bibinfo
  {journal} {Phys. Rev. B}\ }\textbf {\bibinfo {volume} {67}},\ \bibinfo
  {pages} {104401} (\bibinfo {year} {2003})}\BibitemShut {NoStop}%
\bibitem [{\citenamefont {Verissimo}\ \emph {et~al.}(2021)\citenamefont
  {Verissimo}, \citenamefont {Pereira},\ and\ \citenamefont {Lyra}}]{LyraPRB}%
  \BibitemOpen
  \bibfield  {author} {\bibinfo {author} {\bibfnamefont {L.~M.}\ \bibnamefont
  {Verissimo}}, \bibinfo {author} {\bibfnamefont {M.~S.~S.}\ \bibnamefont
  {Pereira}}, \ and\ \bibinfo {author} {\bibfnamefont {M.~L.}\ \bibnamefont
  {Lyra}},\ }\href {https://doi.org/10.1103/PhysRevB.104.024409} {\bibfield
  {journal} {\bibinfo  {journal} {Phys. Rev. B}\ }\textbf {\bibinfo {volume}
  {104}},\ \bibinfo {pages} {024409} (\bibinfo {year} {2021})}\BibitemShut
  {NoStop}%
\bibitem [{\citenamefont {Nico-Katz}\ and\ \citenamefont
  {Bose}(2023)}]{arXiv2207_05052}%
  \BibitemOpen
  \bibfield  {author} {\bibinfo {author} {\bibfnamefont {A.}~\bibnamefont
  {Nico-Katz}}\ and\ \bibinfo {author} {\bibfnamefont {S.}~\bibnamefont
  {Bose}},\ }\href {\doibase 10.1103/PhysRevResearch.5.013041} {\bibfield
  {journal} {\bibinfo  {journal} {Phys. Rev. Research}\ }\textbf {\bibinfo
  {volume} {5}},\ \bibinfo {pages} {013041} (\bibinfo {year}
  {2023})}\BibitemShut {NoStop}%
\bibitem [{\citenamefont {White}(1992)}]{White1}%
  \BibitemOpen
  \bibfield  {author} {\bibinfo {author} {\bibfnamefont {S.~R.}\ \bibnamefont
  {White}},\ }\href {\doibase 10.1103/PhysRevLett.69.2863} {\bibfield
  {journal} {\bibinfo  {journal} {Phys. Rev. Lett.}\ }\textbf {\bibinfo
  {volume} {69}},\ \bibinfo {pages} {2863} (\bibinfo {year}
  {1992})}\BibitemShut {NoStop}%
\bibitem [{\citenamefont {White}(1993)}]{White2}%
  \BibitemOpen
  \bibfield  {author} {\bibinfo {author} {\bibfnamefont {S.~R.}\ \bibnamefont
  {White}},\ }\href {\doibase 10.1103/PhysRevB.48.10345} {\bibfield  {journal}
  {\bibinfo  {journal} {Phys. Rev. B}\ }\textbf {\bibinfo {volume} {48}},\
  \bibinfo {pages} {10345} (\bibinfo {year} {1993})}\BibitemShut {NoStop}%
\bibitem [{\citenamefont {Schollw\"ock}(2005)}]{Schollwock1}%
  \BibitemOpen
  \bibfield  {author} {\bibinfo {author} {\bibfnamefont {U.}~\bibnamefont
  {Schollw\"ock}},\ }\href {\doibase 10.1103/RevModPhys.77.259} {\bibfield
  {journal} {\bibinfo  {journal} {Rev. Mod. Phys.}\ }\textbf {\bibinfo {volume}
  {77}},\ \bibinfo {pages} {259} (\bibinfo {year} {2005})}\BibitemShut
  {NoStop}%
\bibitem [{\citenamefont {Tzeng}(2012)}]{ZengPRB2012}%
  \BibitemOpen
  \bibfield  {author} {\bibinfo {author} {\bibfnamefont {Y.-C.}\ \bibnamefont
  {Tzeng}},\ }\href {\doibase 10.1103/PhysRevB.86.024403} {\bibfield  {journal}
  {\bibinfo  {journal} {Phys. Rev. B}\ }\textbf {\bibinfo {volume} {86}},\
  \bibinfo {pages} {024403} (\bibinfo {year} {2012})}\BibitemShut {NoStop}%
\bibitem [{\citenamefont {Fisher}\ and\ \citenamefont
  {Barber}(1972)}]{FisherPRL}%
  \BibitemOpen
  \bibfield  {author} {\bibinfo {author} {\bibfnamefont {M.~E.}\ \bibnamefont
  {Fisher}}\ and\ \bibinfo {author} {\bibfnamefont {M.~N.}\ \bibnamefont
  {Barber}},\ }\href {\doibase 10.1103/PhysRevLett.28.1516} {\bibfield
  {journal} {\bibinfo  {journal} {Phys. Rev. Lett.}\ }\textbf {\bibinfo
  {volume} {28}},\ \bibinfo {pages} {1516} (\bibinfo {year}
  {1972})}\BibitemShut {NoStop}%
\bibitem [{\citenamefont {{C. Degli Esposti Boschi}}\ and\ \citenamefont {{F.
  Ortolani}}(2004)}]{OrtolaniEPJB}%
  \BibitemOpen
  \bibfield  {author} {\bibinfo {author} {\bibnamefont {{C. Degli Esposti
  Boschi}}}\ and\ \bibinfo {author} {\bibnamefont {{F. Ortolani}}},\ }\href
  {https://epjb.epj.org/articles/epjb/abs/2004/20/b04065/b04065.html}
  {\bibfield  {journal} {\bibinfo  {journal} {Eur. Phys. J. B}\ }\textbf
  {\bibinfo {volume} {41}},\ \bibinfo {pages} {503} (\bibinfo {year}
  {2004})}\BibitemShut {NoStop}%
\bibitem [{\citenamefont {Liu}\ \emph {et~al.}(2022)\citenamefont {Liu},
  \citenamefont {Fu}, \citenamefont {Yu}, \citenamefont {Yu},\ and\
  \citenamefont {Xie}}]{vCTMRG}%
  \BibitemOpen
  \bibfield  {author} {\bibinfo {author} {\bibfnamefont {X.~F.}\ \bibnamefont
  {Liu}}, \bibinfo {author} {\bibfnamefont {Y.~F.}\ \bibnamefont {Fu}},
  \bibinfo {author} {\bibfnamefont {W.~Q.}\ \bibnamefont {Yu}}, \bibinfo
  {author} {\bibfnamefont {J.~F.}\ \bibnamefont {Yu}}, \ and\ \bibinfo {author}
  {\bibfnamefont {Z.~Y.}\ \bibnamefont {Xie}},\ }\href
  {https://cpl.iphy.ac.cn/10.1088/0256-307X/39/6/067502} {\bibfield  {journal}
  {\bibinfo  {journal} {Chin. Phys. Lett.}\ }\textbf {\bibinfo {volume} {39}},\
  \bibinfo {pages} {067502} (\bibinfo {year} {2022})}\BibitemShut {NoStop}%
\bibitem [{\citenamefont {Vidal}(2004)}]{vidal_TEBDPRL2004}%
  \BibitemOpen
  \bibfield  {author} {\bibinfo {author} {\bibfnamefont {G.}~\bibnamefont
  {Vidal}},\ }\href {https://doi.org/10.1103/PhysRevLett.93.040502} {\bibfield
  {journal} {\bibinfo  {journal} {Phys. Rev. Lett.}\ }\textbf {\bibinfo
  {volume} {93}},\ \bibinfo {pages} {040502} (\bibinfo {year}
  {2004})}\BibitemShut {NoStop}%
\bibitem [{\citenamefont {Vidal}(2007)}]{vidal_iTEBDPRL2007}%
  \BibitemOpen
  \bibfield  {author} {\bibinfo {author} {\bibfnamefont {G.}~\bibnamefont
  {Vidal}},\ }\href {https://doi.org/10.1103/PhysRevLett.98.070201} {\bibfield
  {journal} {\bibinfo  {journal} {Phys. Rev. Lett.}\ }\textbf {\bibinfo
  {volume} {98}},\ \bibinfo {pages} {070201} (\bibinfo {year}
  {2007})}\BibitemShut {NoStop}%
\bibitem [{\citenamefont {Vidal}(2008)}]{vidal_iTEBDPRB2008}%
  \BibitemOpen
  \bibfield  {author} {\bibinfo {author} {\bibfnamefont {G.}~\bibnamefont
  {Vidal}},\ }\href {https://doi.org/10.1103/PhysRevB.78.155117} {\bibfield
  {journal} {\bibinfo  {journal} {Phys. Rev. B}\ }\textbf {\bibinfo {volume}
  {78}},\ \bibinfo {pages} {155117} (\bibinfo {year} {2008})}\BibitemShut
  {NoStop}%
\bibitem [{\citenamefont {Blatt}\ \emph {et~al.}(1962)\citenamefont {Blatt},
  \citenamefont {Boer},\ and\ \citenamefont {Brandt}}]{BEC1}%
  \BibitemOpen
  \bibfield  {author} {\bibinfo {author} {\bibfnamefont {J.~M.}\ \bibnamefont
  {Blatt}}, \bibinfo {author} {\bibfnamefont {K.~W.}\ \bibnamefont {Boer}}, \
  and\ \bibinfo {author} {\bibfnamefont {W.}~\bibnamefont {Brandt}},\ }\href
  {https://doi.org/10.1103/PhysRev.126.1691} {\bibfield  {journal} {\bibinfo
  {journal} {Phys. Rev.}\ }\textbf {\bibinfo {volume} {126}},\ \bibinfo {pages}
  {1691} (\bibinfo {year} {1962})}\BibitemShut {NoStop}%
\bibitem [{\citenamefont {Huang}(1987)}]{BEC2}%
  \BibitemOpen
  \bibfield  {author} {\bibinfo {author} {\bibfnamefont {K.}~\bibnamefont
  {Huang}},\ }\href@noop {} {\emph {\bibinfo {title} {Statistical Mechanics}}}\
  (\bibinfo  {publisher} {John Wiley and Sons, New York},\ \bibinfo {year}
  {1987})\BibitemShut {NoStop}%
\bibitem [{\citenamefont {Morita}\ \emph {et~al.}(2022)\citenamefont {Morita},
  \citenamefont {Yoshioka},\ and\ \citenamefont {Kuwata-Gonokami}}]{BEC3}%
  \BibitemOpen
  \bibfield  {author} {\bibinfo {author} {\bibfnamefont {Y.}~\bibnamefont
  {Morita}}, \bibinfo {author} {\bibfnamefont {K.}~\bibnamefont {Yoshioka}}, \
  and\ \bibinfo {author} {\bibfnamefont {M.}~\bibnamefont {Kuwata-Gonokami}},\
  }\href {https://www.nature.com/articles/s41467-022-33103-4} {\bibfield
  {journal} {\bibinfo  {journal} {Nat. Commun.}\ }\textbf {\bibinfo {volume}
  {13}},\ \bibinfo {pages} {5388} (\bibinfo {year} {2022})}\BibitemShut
  {NoStop}%
\end{thebibliography}%

\end{document}